# Non-universal current flow near the metal-insulator transition in an oxide interface

Eylon Persky[1], Naor Vardi[1], Ana Mafalda R.V.L. Monteiro[2], Thierry C. van Thiel[2], Hyeok Yoon[3,4], Yanwu Xie[3,4,5], Benoît Fauqué[6], Andrea D. Caviglia[2], Harold Y. Hwang[3,4], Kamran Behnia[6], Jonathan Ruhman[1], and Beena Kalisky[1]*.

1. Department of Physics and Institute of Nanotechnology and Advanced Materials, Bar-Ilan University, Ramat Gan 5290002, Israel.
2. Kavli Institute of Nanoscience, Delft University of Technology, P.O. Box 5046, 2600 GA Delft, The Netherlands.
3. Geballe Laboratory for Advanced Materials, Department of Applied Physics, Stanford University, Stanford, California 94305, USA.
4. Stanford Institute for Materials and Energy Sciences, SLAC National Accelerator Laboratory, Menlo Park, California 94025, USA.
5. Department of Physics, Zhejiang University, Hangzhou 310027, China.
6. Laboratoire Physique et Etude de Matériaux (CNRS-UPMC), ESPCI Paris, PSL Research University, 75005 Paris, France.

*beena@biu.ac.il

**Abstract**

In systems near phase transitions, macroscopic properties often follow algebraic scaling laws, determined by the dimensionality and the underlying symmetries of the system. The emergence of such universal scaling implies that microscopic details are irrelevant. Here, we locally investigate the scaling properties of the metal-insulator transition at the $LaAlO_3/SrTiO_3$ interface. We show that, by changing the dimensionality and the symmetries of the electronic system, coupling between structural and electronic properties prevents the universal behavior near the transition. By imaging the current flow in the system, we reveal that structural domain boundaries modify the filamentary flow close to the transition point, preventing a fractal with the expected universal dimension from forming. Our results offer a generic platform to engineer electronic transitions on the nanoscale.

**Main Body**

Universal scaling laws in systems near phase transitions are one of the hallmark discoveries of twentieth century physics; near critical points, the thermodynamic properties of fundamentally different systems follow the same algebraic scaling laws[1]. This property allows us to strip complex systems of their microscopic details and characterize them using only their dimensionality and underlying symmetries. For example, at a critical porosity, water percolating through the ground flows along a sub-dimensional fractal, whose dimension is independent of the soil's details[2]. Fractals with the same universal dimension emerge in a variety of other systems, such as forest fires, galactic structures, and metal-insulator transitions[3].

Metal insulator transitions in complex materials, like transition metal oxides, offer an experimental platform for testing the applicability of the universal scaling description. There are clear experimental observations of critical scaling consistent with percolation in a variety of transition metal oxides[4,5]. However, the electronic transition is often intertwined with other material properties, such as magnetic and structural orders[6,7]. Although such couplings can change the dimensionality or symmetries of the electronic systems, it is unclear whether these orders interfere with the expected universal criticality[8–11]. Thus, to determine the mechanism driving the electronic transition, it is essential to resolve the interplay between the electronic and structural orders.

Investigating how domain patterns modify the electronic transition requires local tools, capable of both resolving the patterns, and discerning their effect on electronic properties. We therefore used scanning superconducting quantum interference device (SQUID) microscopy to image the current flow in proximity to the gate-tunable MIT at the $LaAlO_3/SrTiO_3$ (LAO/STO) interface. On the metallic side, the conductivity of LAO/STO is weakly modulated (in space) over structural domain patterns, which emerge from the STO substrate[12–15]. They lead to narrow, elongated, highly conducting channels, with lengths comparable to those of mesoscopic devices[16]. In LAO/STO, there is a gate tunable superconductor-insulator transition[17–19]. At higher temperatures, an MIT has been observed through resistivity[20–22] and compressibility measurements[23]. Resistivity measurements revealed a percolation type transition[19–21], but sample-to-sample variations near the critical point have been reported[20,24]. Here, we reveal that in the presence of domain boundaries, the critical behavior of LAO/STO is not universal. We show that metallicity persists along domain boundaries, as the bulk 2D system turns insulating. As a result, the current carrying backbone cannot scale with the expected universal fractal dimension. Using random resistor network simulations, we show that the lack of a universal backbone coexists with universal scaling of the conductivity, and that the conductivity threshold is size dependent. This combination of universal and non-universal properties suggests that transitions in complex materials must be probed over multiple length scales, to discern their true properties.

To map the current density, we imaged the magnetic fields generated by the current, and used Fourier analysis to reconstruct the current density (Methods and Supplementary Information). To tune the metallicity of the system, we applied a gate voltage ($V_G$) between the 2D interface, and a metallic surface at the bottom of the sample, so that reducing $V_G$ pushed the sample toward its insulating phase. We first show that, in the absence of domain boundaries, the transition is locally consistent with percolation. In a mono-domain, on the metallic side, the current density was overall homogeneous, except for one local reduction due to a patch of reduced conductivity (Figure 1a). Disorder increased as we lowered $V_G$: more insulating patches appeared, and existing patches expanded (Figure 1b, c). Particularly, previously disjoint patches merged to form large insulating regions, limiting the current flow to paths that avoid the insulating regions. The gradual increase in inhomogeneity is consistent with previous transport studies[19–21], which suggest the MIT in LAO/STO occurs via percolation. There are two spatially separated phases – metallic and insulating, and the transition occurs when the metallic phase percolates through the insulating one.

Next, we show that domain boundaries prevent the current carrying backbone from forming a fractal. Figure 2a shows the current flow in a mesoscopic device, with domain boundaries oriented perpendicular to the overall direction of the current flow. Current flow along channels with such orientation is unfavorable, and therefore, the amount of current found on these channels measures the level of inhomogeneity in the surrounding mono-domain regions (Supplementary Information). Indeed, as we reduced $V_G$, the current along the boundaries increased together with the disorder in the mono-domains. At the lowest $V_G$, the current focused along complicated, curved paths in the mono-domain regions, interrupted by straight lines along the boundaries. The current flow along the boundaries is inconsistent with the expected fractal structure: fractals scale sub-dimensionally with the system size ($d_f$ = 1.4 for 2D percolation)[3], which is inconsistent with current paths that are one dimensional over a finite region. Thus, in a multi-domain system, the current carrying backbone cannot scale with the correct universal exponent.

Next, we consider the statistical distribution of the current density. Figure 2b shows logarithmically binned histograms of the current densities, in the metallic phase ($V_G$ = 90 V) and close to the transition ($V_G$ = -66 V). The histograms transition from a narrow distribution, consistent with homogeneous conductivity in the metallic phase, to a wide distribution at $V_G$ = -66 V, which spans three decades of current density. We also considered the current distributions in different mono-domains, compared with that of the overall image. At $V_G$ = -66 V, histograms of different mono-domains show large variations with respect to the histogram of the overall image, particularly in the width of the distribution (Figure 2c). Because the metallic boundaries separate the system into smaller, independent regions, the variations in the histograms may be attributed to a varying disorder landscape in the system. The relevant length-scale for percolation in the mono-domains is set by the distance between domain boundaries, not the overall system size. Thus, the critical behavior is detail dependent, rather than universal.

Even though the current-carrying backbone we find here does not have the universal form expected from a percolation transition, previous resistivity measurements yielded the expected universal critical exponents[19–21]. To see how a non-universal backbone can coexist with universal scaling of the resistivity, we used random resistor network (RRN) simulations. The simulations enable us to study how the critical behavior depends on the system size, a parameter not easily tuned in experiments. We show that, although conductivity can be rescaled with the correct critical exponents, the critical concentration strongly depends on the exact domain pattern, and the overall sample size. As a result, the transition is suppressed for macroscopic systems, and the critical scaling only occurs as a finite-size effect.

We begin by considering the scaling properties of a mono-domain. To model mono-domains, we used a standard network on a 2D square lattice (Figure 3a, Methods). Each node in the network is either metallic (with probability p) or insulating (with probability 1-p). Although the RRN treatment simplifies the transport properties of LAO/STO, it enables us to study the universal scaling behavior near the critical point; similar models were previously used to describe the superconductor-metal-insulator transition in LAO/STO mono-domains[25,26]. This mono-domain network has a phase transition when the concentration of metallic nodes, p, is p = $p_c$ = 0.5[3,27]. At $p_c$, current only flows through a small portion of the metallic bonds, forming the expected sub-dimensional fractal. Correspondingly, the global conductivity of the network shows clear data collapse (Figure 3d,e) when the data are rescaled according to[27]

$$\frac{\sigma}{h^s} = \varphi\left(\frac{p - p_c}{h^{t/s}}\right),$$

where $h = \sigma_I/\sigma_M$ is ratio between the conductivities of metallic and insulating bonds, $\varphi$ is a universal scaling function, and $s = 0.5$ and $t = 1.3$ are the universal scaling exponents for percolation on a 2D square lattice.

We included domain boundaries by modifying the RRN. We considered a set of boundaries oriented perpendicular to the overall direction of current flow, to mimic the experimental data (Figure 2d-g). We modeled the boundary as a set of highly conductive nodes, with conductivity $\sigma_B > \sigma_M$. In this model, we simplified the transport on the highly conducting channels by assuming it is independent of the gate voltage (metallic fraction), and that the domain structure does not change with the applied gate. We neglected gate induced domain motion because we did not observe it in the samples studied. Domain boundaries in LAO/STO are expected to host 1D wave guides with enhanced carrier density[28,29]. We expect the detailed electronic properties to change the conductivity of the insulating phase. However, the current flow images show that the channels remain highly conductive even as the bulk turns insulating. It is therefore appropriate to neglect the detailed electronic structure when considering the effects on critical behavior. Comparing the current density maps at the presumed critical point, p = 0.5, confirms the experimental observation, that the current carrying backbone contains straight narrow lines, inconsistent with a fractal structure. As a result, the conductivity data no longer collapse when rescaled with p$_c$ = 0.5 (Figure 3i). The data collapse is recovered at a lower concentration (Figure 3j). Thus, the introduction of domain boundaries pushes the critical point to a lower value. Recovering the universal scaling is possible because the overall conductivity of the system is determined by the current flow in the mono-domains, rather than the boundaries. Thus, measuring the conductivity probes the mono-domains, which do abide by the expected scaling laws.

Next, we show that the critical concentration depends on the system size, suggesting that the boundaries suppress the transition in the thermodynamic (large system) limit. We simulated additional RRNs, with various system sizes and domain patterns (Figure 3k, Supplementary Information). Domain patterns in mesoscopic LAO/STO devices, often consist of long boundaries, arranged in comb-like patterns[16]. Thus, to describe large systems, we considered uniformly spaced boundaries, whose length is the system size. For each network, we found the critical concentration which recovered the scaling behavior. The resulting critical concentrations (Figure 3k, Supplementary Information) depend on the orientation and spacing of the domain boundaries, and, crucially, on the overall network size. As the network size increases, the critical point is pushed to lower values, suggesting that there is no transition for large networks.

To understand the size dependent shift, we note that, in a standard RRN, the probability to percolate from one side of the sample to the other is given by $P \sim D\, e^{-D/\xi}$, where D is the lateral size of the sample and $\xi \sim |p - p_c|^{-\nu}$ is the correlation length (for 2D percolation, $\nu = 4/3$). In the thermodynamic limit ($D \to \infty$), this probability vanishes, and the sample becomes insulating. When domain walls are present, the largest distance between walls, $l_{\max}$, controls the percolation probability, $P \sim De^{-l_{\max}/\xi}$, which is no longer exponentially small in D. As a result, the critical probability depends logarithmically on D (Supplementary Information), leading to the breakdown of the universal scaling form of Equation 1. Thus, we conclude that the transition in mesoscopic systems containing domain boundaries cannot be described in terms of finite-size scaling.

The simulations and experimental data therefore reveal a surprising combination of critical behaviors. In finite samples, local features (such as the current carrying backbone) do not show the expected universal structure, whereas global features such as the conductivity scale correctly, at a size-dependent critical concentration, because the conductivity is determined by the largest mono-domain region. These results suggest that sample-to-sample variations observed in LAO/STO samples at low carrier densities (see Supplementary Information for discussion) can be explained by variations in the domain patterns. Thus, studying the critical properties of mesoscopic systems requires probing them both on local and global scales.

The lack of a fractal backbone also offers an opportunity to use the gate tunable MIT to study the electronic properties of individual domain boundaries. For example, orienting boundaries along the overall current flow could generate a 2D to 1D crossover, as the current will flow entirely along the boundaries. In STO, this is a particularly promising application, as domain boundaries support a variety of properties, not available in the bulk: they serve as 1D ballistic wave guides[30–32] and host 1D superconductivity[28], electric polarity[13], and magnetic properties[33]. While the realization of such 1D devices requires better control over the size and orientations of domain boundaries, our results show how to decouple the 1D nanowire from the host 2D system.

To conclude, we have shown that, near the MIT in LAO/STO, domain boundaries prevent onset of universal critical behavior. The boundaries hinder the formation of a universal fractal current carrying backbone, and modify the scaling behavior of the conductivity, by making the critical point size dependent. Our results suggest that when electronic properties are coupled to other orders, critical behavior depends on the finite-sized domain patterns, making it non-universal.

Our results also offer a generic platform for engineering the current flow in other oxide heterostructures. Through interface engineering, the domain patterns of the substrate can be imprinted onto the structure of thin films. Coupling between carriers in thin films and the structural properties or phonon modes of their substrates have been reported in a variety of heterostructures[42–45]. Thus, engineering these couplings can generally lead to nanoscale control over the electronic properties and critical behavior of the films.

Coupling between electronic and lattice structures is characteristic of transition metal oxides[6,7]; MITs often occur together with a magnetic or structural phase transition[8,11,37–39,46–48]. To understand the electronic transitions, it is essential to determine how the structural/magnetic transitions contribute, and whether they are the driving force behind the MIT. These questions remain under debate[46,47,49], because the structural and electronic properties are difficult to untangle. Our results suggest that, by modifying the filamentary flow, coupling between structural and electronic properties can create novel critical behavior. For example, near a structural transition, a metallic filament consisting of domain boundaries should be extremely susceptible to changes and can therefore strongly perturb the electronic transition.

Finally, we come back to the relationship between algebraic scaling and fractals. We showed that scaling exponents of global properties do not determine the fractal dimension on the microscopic scale. Our results therefore provide a counterexample to the hyper-scaling relations, suggesting that algebraic scaling of macroscopic quantities does not always imply self-similarity on the microscopic scale.

**Methods**

LAO films were deposited on a TiO$_2$ terminated STO substrate, using pulsed laser deposition[50]. Growth conditions are described in the Supplementary Information. The samples were patterned using an AlO$_x$ hard mask[51]. Electrical contacts to the sample by ultrasonic wire bonding directly to the interface. Alternating currents (1 μA - 60 μA rms, with frequencies between 200 Hz and 2 kHz) were applied to the sample. The resulting magnetic flux was recorded with a scanning SQUID with a 1 μm pickup coil, using a lock-in amplifier. The current density was reconstructed using Fourier analysis[52], and raw magnetic flux data are shown in the Supplementary Information. Data shown in Figure 1 were taken at 4.2 K and data shown in Figure 2 were taken at 5.5 K. The gate leakage current was lower than 1 nA in all measurements.

RRN simulations were performed on square lattice grids with dimensions D×D. Each node is assigned one of three labels: metal, insulator, or domain boundary. Metal and insulator labels were assigned randomly. Metallic nodes were given conductivity $\sigma_M = 1$, domain walls were given conductivity $\sigma_B = 5$, and the conductivity of insulating nodes was varied between simulations. Kirchhoff's rules were then applied for each site, to obtain a set of linear equations for the voltage drop on each resistor. The boundary conditions were set so that current is sourced on one side of the square and drained on the opposite site. The equations were solved via matrix inversion, after one equation had been eliminated, fixing the overall potential shift. The current on each node was then inferred using Ohm's law. The critical point was obtained by minimizing the distance between the rescaled curves (Equation 1), using p$_c$ as the only fitting parameter. Concentrations with $|p - p_c| < D^{-1/\nu}$ were not used for the fit, to avoid effects stemming from the finite dimensions of the network. Error bars were obtained by statistical bootstrapping.

**Acknowledgements**

We thank Herb A. Fertig, Ganapathy Muthry, Efrat Shimshoni and Brian Skinner for fruitful discussions. E.P., N.V., and B.K. were supported by European Research Council Grant No. ERC-2019-COG-866236, and Israeli Science Foundation grant no. ISF-1281/17. J.R. was supported by Israeli Science Foundation grant no. 967/19. A.D.C. was supported by European Research Council Grant No. ERC-2015-STG-677458, and by The Netherlands Organisation for Scientific Research (NWO/OCW) as part of the VIDI programme. B.K., and A.D.C. were supported by the QuantERA ERA-NET Cofund in Quantum Technologies (Project No. 731473). Work at Stanford was supported by the Department of Energy, Office of Basic Energy Sciences, Division of Materials Sciences and Engineering, under contract No. DE-AC02-76SF00515.

**Author contributions**

E.P. and B.K. conceived the work and designed the experiments. E.P, N.V. and B.K. performed the measurements. A.M.R.V.L.M., T.C.T., H.Y., Y.X., A.D.C. and H.Y.H. fabricated the samples. E.P. and J.R. performed the RRN modeling. E.P., B.K., J.R, A.D.C., B.F. and K.B. interpreted the results. E.P, B.K. and J.R. wrote the manuscript, with contributions from all co-authors.

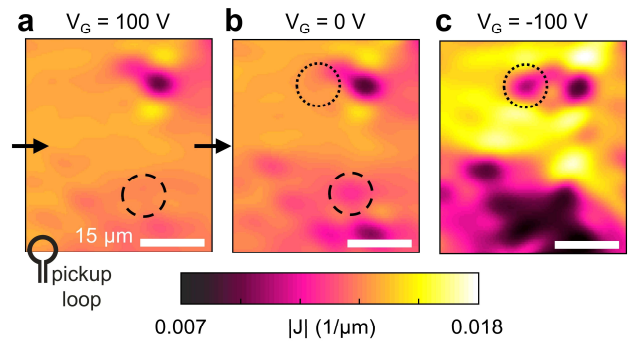

**Figure 1. Disorder and percolation in a mono-domain.** Current density (magnitude) images of a mono-domain region, as a function of $V_G$. Reducing $V_G$ increases the inhomogeneity of the current flow, as additional insulating patches (circles) modify the flow, and isolated insulating patches cluster together (panel c). The arrows in panel a mark the overall direction of the current flow. The SQUID's pickup loop, (diameter 1 μm), is not drawn to scale.

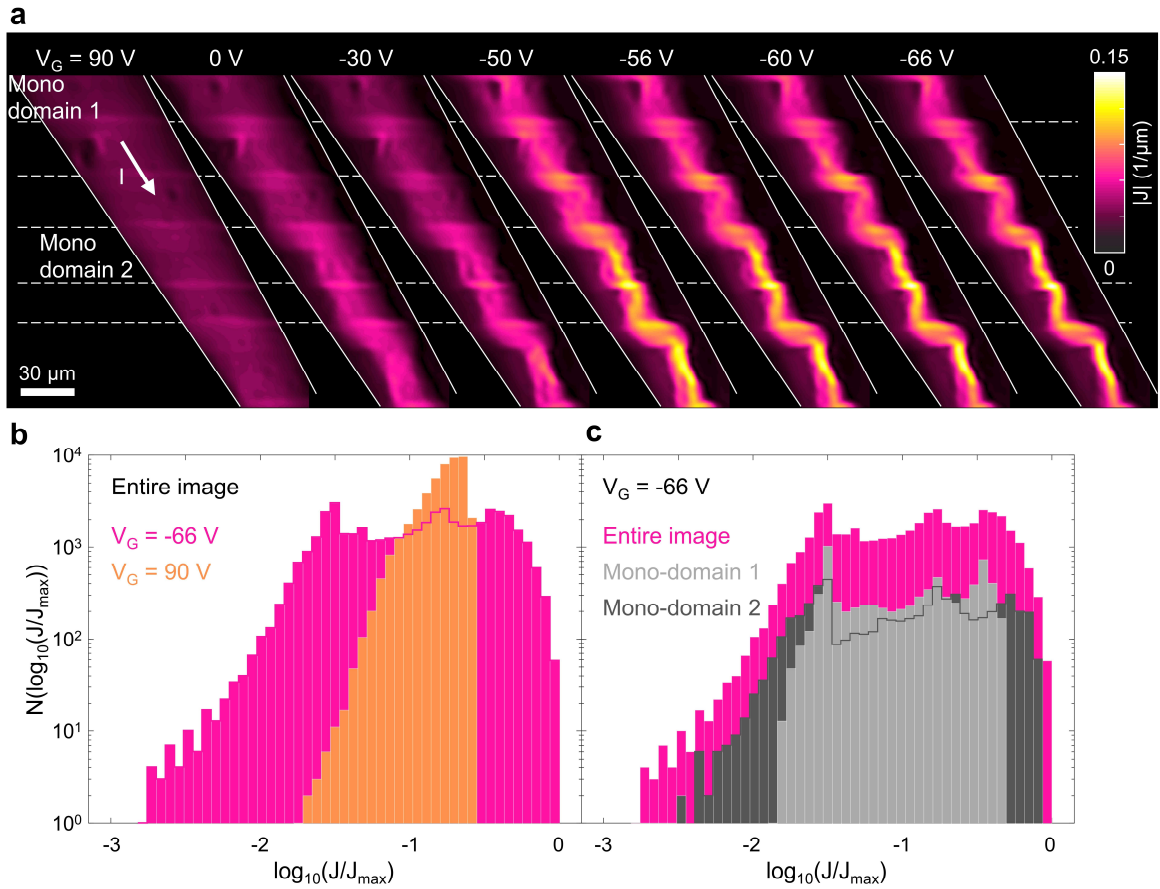

**Figure 2. Non-universal current backbone in a multi-domain device. a**, Current density images of a multi-domain region, as a function of $V_G$. Reducing $V_G$, mono-domain regions became more inhomogeneous, while the boundaries remained conducting. As a result, more current flowed along the boundaries. The rightmost image shows complicated current paths in the mono-domain regions, interrupted by the boundaries. In such patterns the current-carrying backbone cannot have the universal fractal scaling dimension. **b**, Logarithmically binned histogram of the current density of the leftmost and rightmost maps in a, transitioning from highly homogeneous conductivity on the metallic side, characterized by a narrow current distribution to disordered flow closer to the MIT, where the histograms spans three decades of current density. **c**, Histograms of two mono-domain regions at $V_G = -66$ V, showing variations in the current distributions between different mono-domains, particularly in the width of the distributions, and in the low-current behavior.

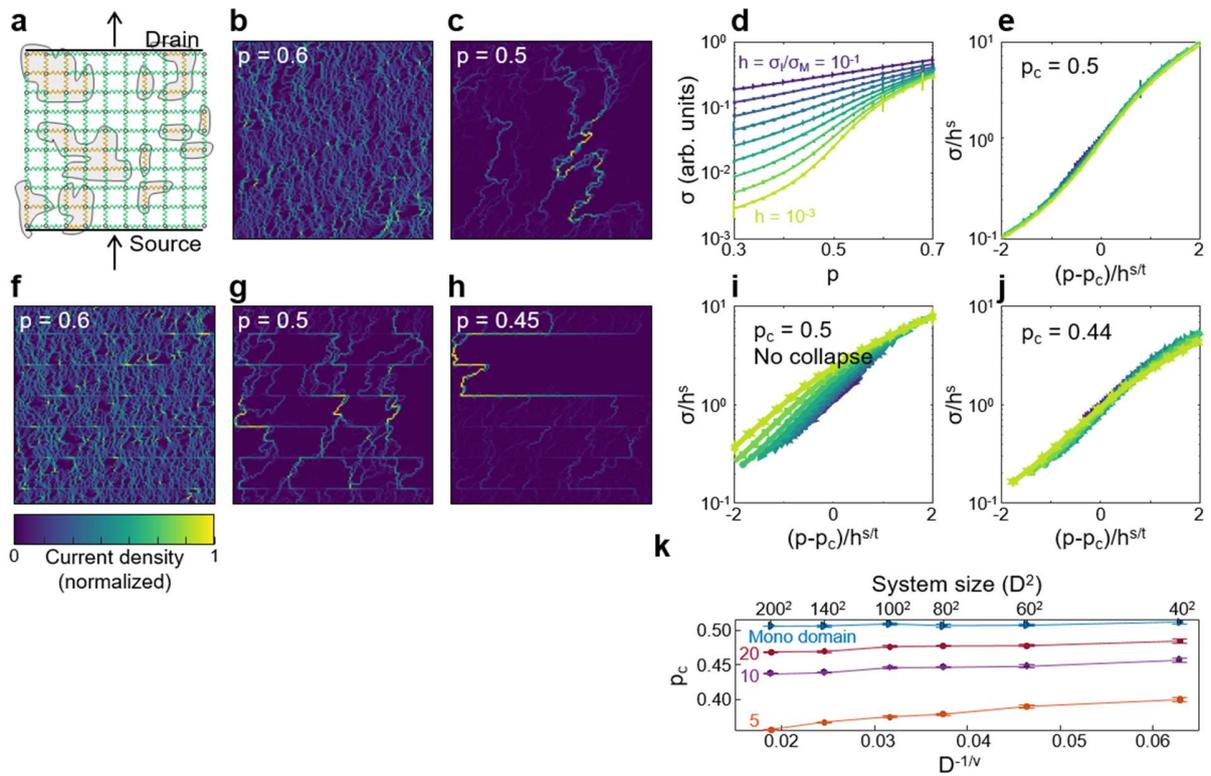

**Figure 3. Size dependence of the metal-insulator transition. a**, Illustration of the random resistor network, showing isolated insulating patches (yellow nodes) within the metallic (green nodes) network. **b,c**, Current density maps obtained from individual realizations of a 128*128 network, at p = 0.6 (b) and p = 0.5 (c). The current was sourced at the bottom side of the square, and drained at the top. At p = 0.5, the current flows along the expected sub dimensional fractal. **d**, The conductivity of the network, plotted as a function of p, for various bond conductivity ratios h. **e**, The data in d rescaled with the universal critical exponents s = 0.5 and t = 1.3, showing data collapse at the critical concentration, $p_c$ = 0.5. **f-h**, Representative realizations of the model with domain boundaries, for p = 0.6 (f), p = 0.5 (g), and p = 0.45 (h). The current carrying backbone in g no longer consists of a single complicated path. Instead, shorter complex paths appear at a lower concentration (h), interconnecting the domain boundaries. The boundaries introduce long, straight lines into the backbone, inconsistent with the expected size scaling of a fractal. **i,j**, Rescaled conductivity data for the critical concentration of the standard model, pc = 0.5 (i), and for the shifted critical point pc = 0.44 (j). **k**, Size dependence of the critical concentration, for networks with various domain wall spacings. The mono-domain network shows no size dependence, while the critical concentrations of multi-domain RRNs is lower for larger systems sizes. The size and detail dependence suggest that the transition is not universal. The error bars represent two standard deviations from the mean fitted critical point.

# Supplementary Information for
# Non-universal current flow near the metal-insulator transition in an oxide interface

Eylon Persky[1], Naor Vardi[1], Ana Mafalda R.V.L. Monteiro[2], Thierry C. van Thiel[2], Hyeok Yoon[3,4], Yanwu Xie[3,4,5], Benoît Fauqué[6], Andrea D. Caviglia[2], Harold Y. Hwang[3,4], Kamran Behnia[6], Jonathan Ruhman[1], and Beena Kalisky[1]*.

7.  Department of Physics and Institute of Nanotechnology and Advanced Materials, Bar-Ilan University, Ramat Gan 5290002, Israel.
8.  Kavli Institute of Nanoscience, Delft University of Technology, P.O. Box 5046, 2600 GA Delft, The Netherlands.
9.  Geballe Laboratory for Advanced Materials, Department of Applied Physics, Stanford University, Stanford, California 94305, USA.
10. Stanford Institute for Materials and Energy Sciences, SLAC National Accelerator Laboratory, Menlo Park, California 94025, USA.
11. Department of Physics, Zhejiang University, Hangzhou 310027, China.
12. Laboratoire Physique et Etude de Matériaux (CNRS-UPMC), ESPCI Paris, PSL Research University, 75005 Paris, France.

*beena@biu.ac.il

**Current reconstruction**

For a two dimensional (2D) current distribution, it is possible to reconstruct the current density, $\boldsymbol{J}(x,y) = J_x(x,y)\,\hat{\boldsymbol{x}} + J_y(x,y)\,\hat{\boldsymbol{y}}$, from the magnetic flux data performed by the SQUID. The magnetic field generated by a current density $\boldsymbol{J}(x,y)$ is given by the Biot-Savart law,

$$B_z(\boldsymbol{r}) = \frac{\mu_0 d}{4\pi} \int dx' \int dy' \frac{\boldsymbol{J}(\boldsymbol{r}') \times (\boldsymbol{r}-\boldsymbol{r}')}{|\boldsymbol{r}-\boldsymbol{r}'|^3} \cdot \hat{\boldsymbol{z}}, \qquad (1)$$

where $B_z$ is the out-of-plane ($\hat{\boldsymbol{z}}$) component of the magnetic field, and $d$ is the thickness of the conducting layer. Interpreting equation (1) as a convolution integral, we use Fourier analysis to deconvolve the kernel function from the magnetic field measurements, recovering the current distribution. Further details about implementation of this technique to scanning SQUID data appear elsewhere[1,2]. Figure S1 demonstrates how the qualitative features of the data can be clearly observed in the magnetic flux (raw) data.

Figure S1a shows a raw magnetic flux image of a representative LAO/STO device. The raw data feature long-scale field gradients, in accordance with the Biot-Savart law for a homogeneous current flow, as well as shorter-scale modulations, corresponding to local changes in the current density. As an alternative way to visualize the local modulations, Figure S1b shows the magnetic flux data, after the application of a high-pass filter (allowing spatial frequencies ≥ 0.4 μm$^{-1}$). The resulting image is proportional to the gradient of the current density, clearly identifying the edges of the device and the local modulations due to defects, while suppressing the homogeneous part of the current flow. Dipole-like features in the filtered image correspond to local reductions of the current density: the current avoids an area with reduced conductivity, increasing the flow around it. The elongated modulations correspond to stripes with enhanced current flow. The same features are directly visible in the current density magnitude image (Figure S1c). The line cuts in Figure S1d demonstrate how locally reduced, locally enhanced, and homogeneous current flow appear in the magnetic flux data, and how the current reconstruction quantifies this effect.

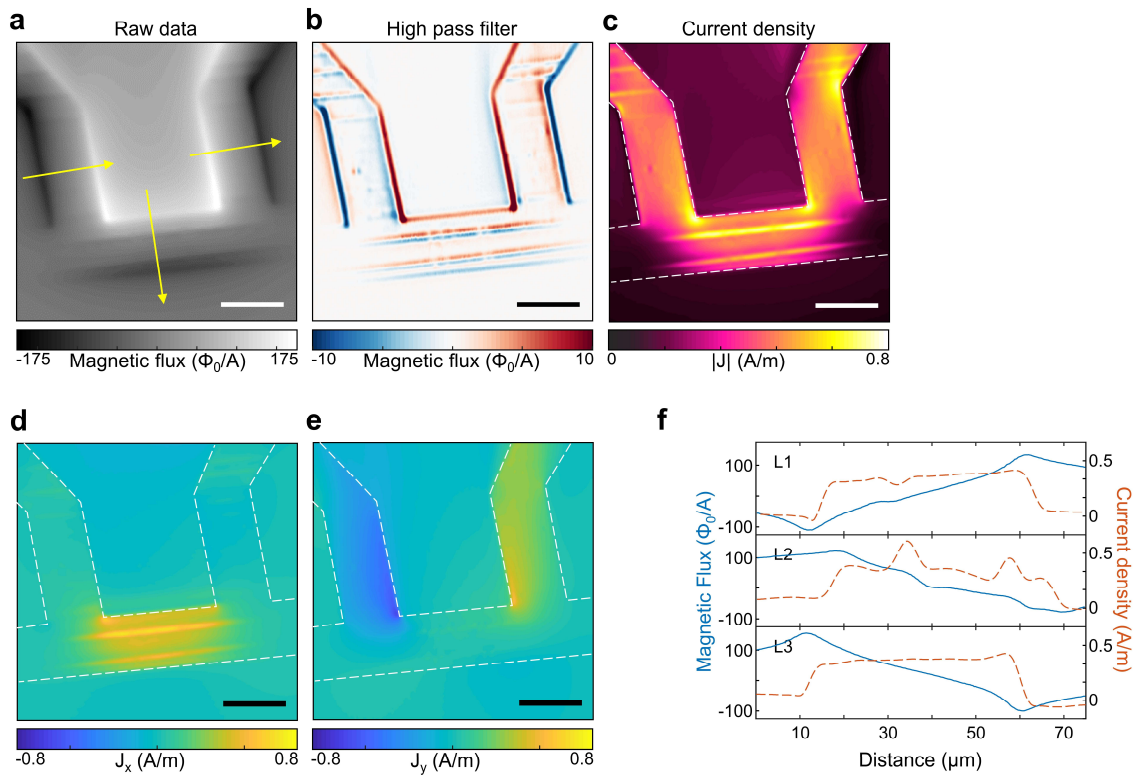

**Figure S1. Current reconstruction. a,** Raw magnetic flux data corresponding to the current density shown in Figure 1c of the main text, due to a current of 20 µA flowing in a device 50 µm wide. **b,** Magnetic flux data after application of a high-pass filter, enhancing the local modulations of the current flow. **c,** The reconstructed current density (magnitude) map. **d, e** Current density components in the x (d) and y (e) directions. Scale bars, 50 µm. **f,** Line cuts taken along the yellow lines in a, showing the magnetic flux profile (solid lines) and the reconstructed current density (dashed lines). Lines L1 and L2 correspond to local reduction and local enhancement of current flow, respectively, while line L3 corresponds to a homogeneous current distribution.

**Distribution of domain boundaries**

Figure S1 demonstrates how the domain wall distribution can vary significantly within a single device: the images reveal large mono-domain areas, neighboring regions with a higher density of walls. This is a common feature in LAO/STO devices. Figure S2a shows another example of the large spatial variations in the domain patterns. Within a single device, (field of view: 250 µm × 250 µm), we observed regions with a low density of boundaries, oriented along the [010] direction (rectangle b), large mono-domain regions (rectangle c), high density of [010] boundaries (rectangle d), and [100] boundaries (rectangle e). The resulting response to the gate, shown in Figures S2b-e shows the range of behaviors discussed in the main text: current focusing along domain walls oriented parallel to the overall direction of the flow, percolation in mono-domain regions, and percolation between domain walls oriented perpendicular to the overall direction of the flow.

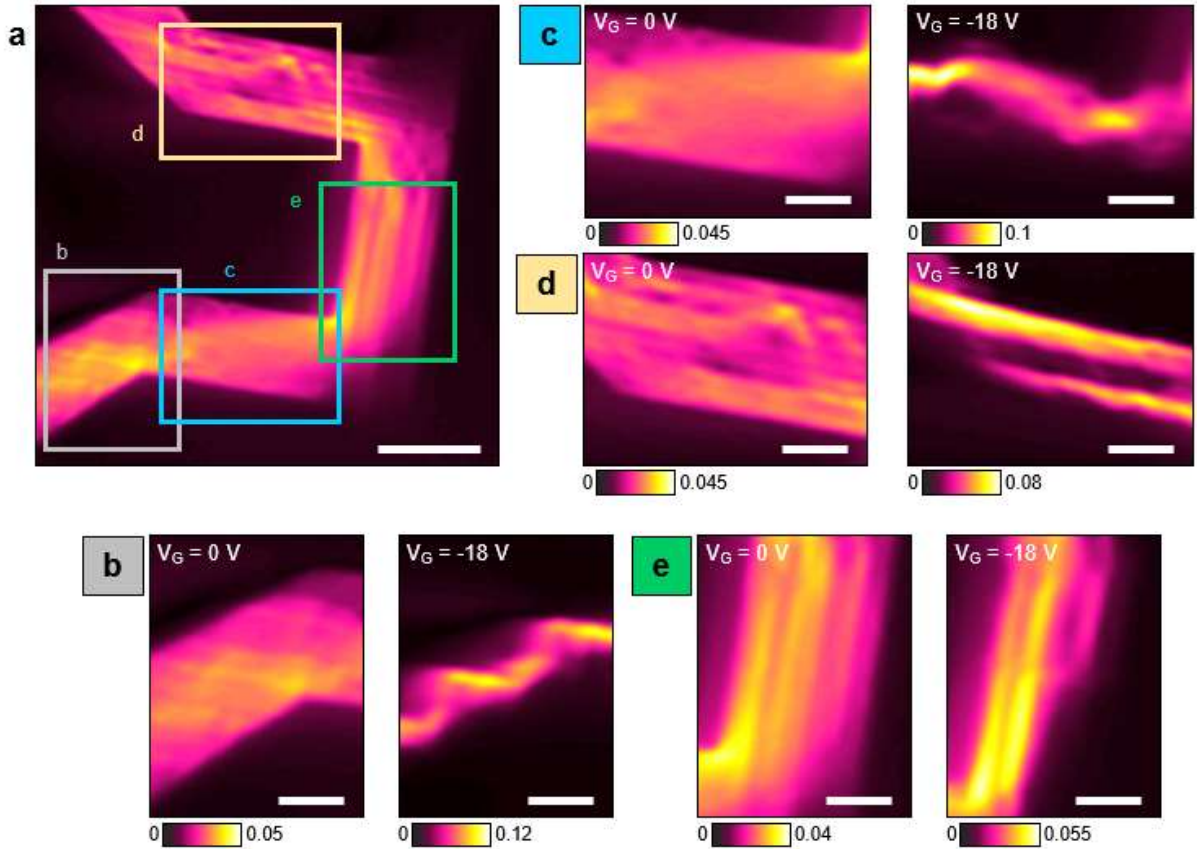

**Figure S2. Current paths near the MIT for various domain patterns on a single device. a,** Large area current density image of a device, showing a various domain patterns. Scale bar, 50 μm. **b-e,** current density maps of the rectangles shown on panel a, at $V_G$ = 0 V (left) and $V_G$ = -18 V (right), showing the different current paths near the MIT, due to the different domain patterns. Scale bars, 25 μm.

**Estimation of the gas parameter**

Here, we estimate the gas parameter, $r_s$, of LAO/STO, quantifying the ratio between the potential and kinetic energy in the system. For a 2D system, $r_s$ is given by $r_s = 1/\sqrt{\pi n (a_B^*)^2}$, where $n$ is the charge carrier density, and $a_B^* = \hbar^2 \epsilon / m^* e^2$ is the effective Bohr radius. The carrier density[3] (~1×10$^{13}$ cm$^{-2}$) and the effective mass[4,5] (~3m$_e$) have been measured. We are left with estimating the dielectric constant $\epsilon$, which controls the screening of the electron-electron interactions. Although the dielectric constant of bulk STO at low temperatures is extremely large[6] (~2×10$^4$), its value close to the 2DEG is significantly lower, since the large electric field generated by the polar LAO layer strongly pins the optical phonon distortion[7]. As a result, an estimate of a surface dielectric constant $\epsilon$~100 is appropriate[8]. The resulting Bohr radius is $a_B^* \cong 2\ nm$, leading to $r_s \cong 1$.

**RRN results for complex domain patterns**

In unpatterned devices, structural domains tend to form strongly correlated patterns, due to the long ranged strain fields they induce. In the main text, we investigated through experiment and simulations an example of one such pattern: an array of elongated, equally spaced domain walls. Here, we used the RRN simulations to consider more complex patterns, which had been observed in previous optical studies[9]. These patterns contain arrays of intersecting boundaries of various orientations and irregular length and spacing (Figure S3a-c). For all patterns we studied, the critical point was shifted downwards with increasing network size, but the rate of change varied considerably between the patterns (Figure S3g).

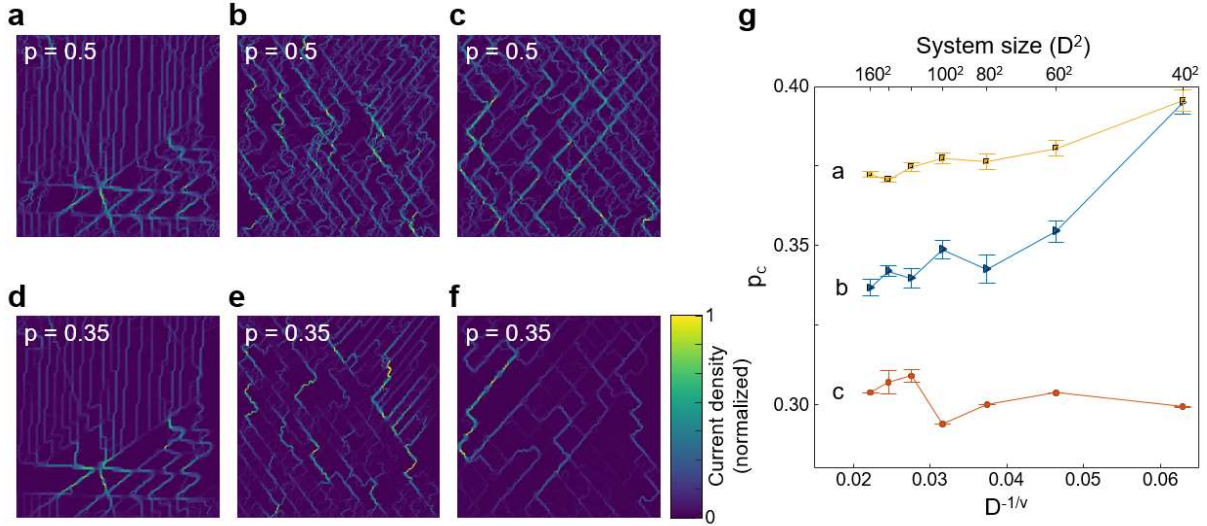

**Figure 3. RRN simulations for complex domain patterns. a-c**, current distributions from single realizations of the models, for various domain patterns which appeared in unpatented samples[9], at p = 0.5. **d-f**, same as panels a-c, but for p = 0.35. **g**, Size dependence of the shifted critical point for each configuration.

**Estimation of the shift to the critical point**

Here, we provide a scaling argument which explains the size-dependent shift of the critical point, due to the introduction of conducting channels. We first consider the case of domain boundaries whose position and size is randomly distributed, with some typical separation $l$. In this case, the transition occurs when the correlation length is of the order $l$,

$$\xi = |p - p_c|^{-\nu} \cong l, \qquad (2)$$

suggesting a shift to the critical point

$$p = p_c - (1/l)^{1/\nu}, \qquad (3)$$

which depends on the spacing $l$ of the walls. Experimentally, such domain wall distributions can be realized in magnetic materials, where an external magnetic field can be used to change the typical size of domains. The distribution of structural domain walls, however, is often strongly correlated. These walls tend to form highly correlated distributions, which are elongated along the crystallographic direction. For this type of distributions, the critical concentration strongly depends on the details of the domain pattern. If the walls are parallel to the overall direction of the current flow, the transition is suppressed altogether. On the other hand, for equally spaced walls perpendicular to the flow, the , the critical value of $p_c$ is pushed to a size-dependent value, $p_{DW}$: when the length of the walls is comparable to the system size, D, the probability to find a percolating path, $P \sim D e^{-l/\xi}$, is no longer exponentially small in D. The transition is therefore shifted to a lower value, logarithmically dependent on D,

$$p_{DW}(D,l) \sim p_c - \left[\frac{\log D}{l}\right]^{\frac{1}{\nu}}. \tag{4}$$

**Domain boundaries perpendicular to the overall direction of the current**

In the metallic regime, the geometry of the device controls the direction of the current flow. The perturbation from domain boundaries depends on their orientation with respect to the device geometry: boundaries oriented along the device (such as those in Figure S2d,e) can draw larger portions of the current than boundaries oriented perpendicular to the overall direction of the current (such as Figure S2b).

We explain this behavior in terms of random resistor networks. First, consider a network with homogeneous conductivity (p=1), where the voltage gradient is applied in the y direction. In this case, the least resistive path between a point at the bottom edge, and the top edge of the network, is the same as the shortest path. The shortest path is a straight line in the y direction, so no current will flow in the x direction. Adding a line of less resistive nodes in the x direction does not change this behavior, because flow in the x direction elongates the path, making it more resistive.

At a finite disorder (p<1), the least resistance path is no longer the shortest path. The path becomes longer because it is favorable to avoid highly resistive nodes. Flow along a perpendicular conductive channel becomes more likely, because nodes immediately before or after the channel may be insulating. Close to the percolation threshold, the modulation is largest, because the highly conducting channels have a high probability of connecting dangling bonds (parts of the percolating cluster with a dead end that does not connect to the electrodes) to the backbone. This significantly modifies the current flow, because it generates several paths connected in parallel, as opposed to a single series path.

**Survey of observed variations in transport at low carrier concentrations**

The non-universal current backbone, and size and geometry dependence of the critical point imply that transport behavior of mesoscopic samples near the MIT must show strong variations between samples. There is a large volume of work gate-tunable properties of LAO/STO, which supports this conclusion. The percolation-type MIT observed through conductivity measurements [3,10] showed large variations in the critical carrier density between different samples[3], and quadratic magneto-resistance near the transition, inconsistent with the linear behavior expected for percolation in an "ordinary" disordered system[10]. Further, there are multiple conflicting reports about the gate tunable magnetoresistance and Shubnikov-de-Hass oscillations (see Ref.[11] for a review). The large variability between studies strongly suggests that the source of inconsistencies is intrinsic to the material system.

Similarly, our results offer a new perspective on the MIT in bulk, three dimensional STO. Even though the experimentally observed Fermi surface of doped crystals[12] is in good agreement with its calculated band structure[13], bulk STO do not display a sharp MIT, and the dilute metal gradually fades away as the doping is reduced[14]. While this idea requires further experimental effort, our results identify domain boundaries as a key player in the emergent metallicity near the putative critical doping, in both bulk STO, and its 2D counterpart.